# Magneto-Transport and High-Resolution Angle-Resolved Photoelectron Spectroscopy Studies of Palladium Doped Bi$_2$Te$_3$

Shailja Sharma[1], Shiv Kumar[2], G.C. Tewari[3], Girish Sharma[1], Eike F. Schwier[2, #], Kenya Shimada[2], A. Taraphder[4], and C.S. Yadav[1]*

[1]*School of Basic Sciences, Indian Institute of Technology Mandi, Mandi-175075 (H.P.) India*
[2]*Hiroshima Synchrotron Radiation Center, Hiroshima University, Hiroshima, Japan*
[3]*Department of Chemistry and Material Science, Aalto University, FI-00076 Aalto, Finland*
[4]*Department of Physics, Indian Institute of Technology Kharagpur, Kharagpur (W.B) India*
*Email: shekhar@iitmandi.ac.in*

We have performed magneto-transport and high-resolution angle-resolved photoelectron spectroscopy (ARPES) measurements on palladium (Pd) doped topological insulator Pd$_x$Bi$_2$Te$_3$ (0 ≤ $x$ ≤ 0.20) single crystals. We have observed unusually high values of magnetoresistance (~1500%) and mobility (~ 93000 cm$^2$V$^{-1}$s$^{-1}$) at low temperatures for pristine Bi$_2$Te$_3$ that decrease on Pd doping. The Shubnikov-de Haas (SdH) oscillations have been detected for $x$ = 0.05, 0.10, confirming the presence of 2D topological surface states (TSSs) for these samples. The Hall measurement shows the crossover from *n*-type charge carriers in pristine Bi$_2$Te$_3$ to *p*-type charge carriers upon Pd doping. The ARPES measurements show that the conduction band crosses the Fermi level for pristine Bi$_2$Te$_3$, and the Dirac point of the TSSs and bulk-derived valence bands indicated shift to lower binding energy upon Pd doping in a rigid-band-like way up to $x$ ~ 0.10. Based on the comparison of the parameters obtained from the SdH and ARPES measurements, the reduction in the $k_F$ value in the magneto-transport measurements likely due to the band bending induced by the Schottky barrier.

## I. INTRODUCTION:

Topological insulators (TIs) are one of the most exciting and studied systems in contemporary condensed matter physics. TIs are characterized by the gapless topological surface state (TSS) which are located inside the bulk band gap. The TSSs remain gapless under the perturbation conserving the topological number which is calculated based on the symmetry properties of the wave functions in the bulk.[1,2] The TSS has a helical spin texture, and its dispersion around the time-reversal invariant momenta (TRIM) can be regarded as that of a massless Dirac fermion.[3] Electrons in the TSS acquires a $\pi$ Berry phase after adiabatically completing a closed trajectory around the Fermi surface.[4] TIs exhibit a variety of exotic electronic transport properties such as non-saturating linear magnetoresistance, low field weak antilocalization effect (WAL), Shubnikov-de Haas (SdH) oscillations, and high carrier mobility.[5] Angle-resolved photoemission spectroscopy (ARPES) and scanning tunneling microscopy studies have directly revealed a Dirac-cone-like TSS in the bulk band gap.[6,7]

The layered Bi$_2$Te$_3$ is one of the typical three-dimensional TIs and its TSSs has been extensively studied through transport and spectroscopy experiments.[6,8] Bi$_2$Te$_3$ crystallizes in rhombohedral crystal structure with space group $R\bar{3}m$ (#166).[9] The unit cell of Bi$_2$Te$_3$ comprises of three quintuple layers (QLs) which are bonded by weak van der Waals forces, and each QL consists of five atomic layers arranged in *Te$^I$-Bi-Te$^{II}$-Bi-Te$^I$* sequence. This weak binding force between Te atoms in adjacent QLs account for the easy cleavage along the *ab*-plane perpendicular to the *c*-axis, and the anisotropic thermal and electronic transport properties arise from these structural properties. The spectroscopy studies indicate that Bi$_2$Te$_3$ is an indirect narrow-band-gap (~150 meV) topological insulator with a single Dirac cone at the $\bar{\Gamma}$ point.[6] Bi$_2$Te$_3$ has been known for the highest thermoelectric figure of merit (*ZT*) around room temperature.[10] Superconductivity has also been observed in Bi$_2$Te$_3$ under high pressures between 3 - 6 GPa with a superconducting transition temperature of $T_c$ ~ 3 K in the rhombohedral phase without structural phase transition.[11-13]

Previously, Hor *et al.*[14] has reported superconductivity at $T_c$ ~ 5.5 K in Pd$_x$Bi$_2$Te$_3$ ($x$ = 0.15, 0.3, 0.5, and 1.0) compounds through ac susceptibility and resistivity measurements. Amit *et al.*[15] discussed the temperature dependent resistivity and magnetic susceptibility at various magnetic fields (*H*) and pressure (*p*) to construct the phase diagram for the superconducting compound PdBi$_2$Te$_3$. The superconducting transition temperature was decreased from $T_c$ ~ 5.5 K with increasing pressure. Both the above reports indicated small superconducting volume fraction ~ 1% for the compound PdBi$_2$Te$_3$, which is likely due to the difficulty in Pd intercalation into the van der Waals gap between the QLs.[14] So far, however, the magneto-transport and high-resolution ARPES measurements of Pd doped Bi$_2$Te$_3$ have not been explored yet. These measurements are quite useful in understanding the electron dynamics and electronic band structure and may give useful information about the topological surface states. Pd is a higher $Z$ element in comparison to other dopants viz. Nb, Cu, and Sr, and known to show exchange enhanced magnetic susceptibility. It is, therefore, worth investigating the Pd doping from the viewpoint of tailoring the electronic transport, spin-orbit coupling, and topological surface states of the $Bi_2Te_3$.

In this study, we present detailed electronic transport properties and high-resolution ARPES studies to investigate the palladium doped bismuth telluride, $Pd_xBi_2Te_3$ single crystals. We have observed a large value of the magnetoresistance, and ultra-high mobility at low temperatures in pristine $Bi_2Te_3$. The evolution of the electronic transport and topological surface state properties upon Pd doping is discussed in the light of magnetoresistance, quantum oscillations, WAL, and the electronic structure revealed by high-resolution ARPES. By combining the results from ARPES and SdH oscillations measurements, we discuss the macroscopic magneto-transport properties from the viewpoint of microscopic electronic band structure.

## II. EXPERIMENTS:

Single crystals of $Pd_xBi_2Te_3$ ($x$ = 0, 0.05, 0.10, 0.15, 0.20) were grown using the flux method in two steps. High purity elemental Bi (99.99%), Te (99.999%), Pd (99.99%) from Sigma-Aldrich Co. were accurately weighed according to the stoichiometric ratios and vacuum sealed (>$10^{-5}$ mbar) in quartz tubes. In the first step, $Bi_2Te_3$ were prepared for all the compositions. The compounds were heated at 850 °C for 24 h, followed by cooling to 500 °C at the rate 10 °C /h, where they were kept for 72 h. Then, the compounds were furnace off cooled to room temperature. Second, for Pd doped samples, the melted $Bi_2Te_3$ and Pd were mixed thoroughly using mortar and pestle, pelletized and vacuum sealed and kept for another heat treatment at 850 °C for 48 h, then slow cooled (3 °C/h) to 400 °C and kept for 12 h. Then, the compounds were furnace off cooled to room temperature. Single crystals were obtained from the ingots easily by cleaving perpendicular to the (*00l*) axis.

The crystal structure and phase purity of the samples were determined by X-ray diffraction measurement at room temperature using Rigaku Smartlab diffractometer with Cu-K$\alpha$ ($\lambda$= 1.5418 Å) radiation. The morphological studies on freshly exfoliated samples were studied using energy dispersive spectroscopy (EDS) and field emission scanning electron microscope (FESEM FEI, Nova Nano SEM 450). Figures S1 and S2 depict the x-ray diffraction patterns, Rietveld refinements and layered morphology of studied samples.[16] The resistivity and Hall effect measurements were performed in a Physical Property Measurement System (PPMS, Quantum Design Inc.) using the standard four-probe contact configuration in the temperature range 1.8 - 300 K up to magnetic fields of 8 T. Magnetization measurement were performed in a Magnetic Property Measurement System (MPMS, Quantum Design Inc.). The magnetic field was perpendicular to the cleaved plane of crystals and the current direction. The samples were exfoliated using scotch tape before performing above measurements.

The high-resolution ARPES measurements were carried out using the μ-Laser ARPES system developed at the Hiroshima Synchrotron Radiation Center, Hiroshima University, Japan.[17] The samples were cleaved *in situ* on the 5 axes manipulator at a temperature of 20 K, using the top-post method. The base pressure was below 5×$10^{-9}$ Pa. The Dirac-cone-like spectra was measured in the *s*-polarization geometry along the $\bar{\Gamma} - \bar{M}$ high symmetry line of the surface Brillouin zone. Azimuthal sample alignment was performed *ex situ* via measurements of Laue patterns, while the $\bar{\Gamma}$-point position was determined by a fine angular map along the direction perpendicular to the dispersive direction of the analyzer slit. The position of the Fermi level was determined using the Fermi edge of the reference sample in electrical contact with the sample. The Laser spot size (on the sample) was less than 10 μm resulting in an intrinsic angular resolution of less than 0.05°. The overall instrumental energy resolution was estimated to be better than 2 meV.

## III. RESULTS:

### A. Transport measurements

Figure 1(a) shows the temperature dependence of longitudinal resistivity ($\rho_{xx}$) of $Pd_xBi_2Te_3$ crystals from 1.8 - 300 K measured under zero field. The $\rho_{xx}$ exhibits metallic behavior throughout the temperature range. The Residual Resistivity Ratio (RRR), which is defined as the ratio of resistivity at room temperature to that at low temperature (RRR = $\rho$(300 K)/$\rho_0$(1.8 K)), decreased on Pd doping from ~ 57 for $Bi_2Te_3$ to ~ 4 for $Pd_{0.20}Bi_2Te_3$. The high value of RRR for $Bi_2Te_3$ is further reflected in the narrow linewidth in high-resolution ARPES spectra which is among the best for the 3D TIs[6,18-24] It has been reported that $Bi_2Te_3$ show distinct resistivity behavior with varying value of RRR for different pieces of the crystal even if they are taken from the same batch grown in the same ampoule.[8] The electronic properties of these compounds strongly depend on the internal defects and inhomogeneity in the material. Theoretical studies by Black-Schaffer *et al.* on the role of subsurface impurities and vacancies in the 3D TIs suggested that bulk impurities can give rise to gapless bulk conductivity and may mask the surface transport properties.[25] Therefore, we measured $\rho_{xx}(T)$ on the four different pieces of the crystal for each composition (See Fig. S3 [16]). Samples taken from some portions of the crystal were found to exhibit a broad hump shape around 100 - 150 K, suggesting that the transport properties are sensitive to the inhomogeneity of the tellurium and anti-site defects present in the crystal. At low temperatures, all the $\rho_{xx}(T)$ show saturation below 10 K, implying the finite residual resistivity. We have observed a signature of superconductivity in $Pd_{0.20}Bi_2Te_3$ below $T_c \sim$ 2.2 K in the magnetization measurements (See M(T) and M(H) curves in Fig. S4[16]). However, we could not detect superconductivity in the resistivity measurements, which is likely due to the low superconducting volume fraction.

Figure 1(b) depicts the variation of charge carrier density (*n*) in the temperature range 1.8 - 300 K. The bulk carrier density was estimated using a single-carrier Drude band model, $n(T) = 1/(eR_H(T))$. The slope has been calculated from the linear part of the magnetic field dependence of $\rho_{xy}$ (See Fig. S5 [16]). It is evident that for all temperatures, $\rho_{xy}$ is linear and negative for $Bi_2Te_3$, implying the presence of *n*-type bulk charge carriers. With an increase in Pd concentration $\rho_{xy}$ changes its sign, manifesting the hole-dominant transport in such systems. We find that the bulk carrier density is in the range $10^{18}$ - $10^{19}$ cm$^{-3}$ for $Pd_xBi_2Te_3$ without any appreciable temperature dependence except for $Pd_{0.05}Bi_2Te_3$. It is to note that there is a sharp change in the carrier concentration of $Pd_{0.05}Bi_2Te_3$ above $T = 20$ K. The transverse resistivity $\rho_{xy}(H)$ of $Pd_{0.05}Bi_2Te_3$ (See Fig. S5 [16]) shows a sudden change in the slope between low temperature and high temperature $\rho_{xy}(H)$ values.

Using Hall coefficient ($R_H$) and longitudinal resistivity ($\rho_{xx}$) data, we have estimated the Hall mobility ($\mu$), which is shown in Fig. 1(c) with the smallest observed carrier density (7.58×10$^{17}$ cm$^{-3}$) and lowest residual resistivity (8.8×10$^{-5}$ Ω-cm) at 1.8 K. These result in ultra-high mobility values of ~ 93000 cm$^2$V$^{-1}$s$^{-1}$ for pristine $Bi_2Te_3$. The extremely large mobility for $Bi_2Te_3$ is in accordance with the high RRR and large MR value (MR varies quadratically with mobility) as discussed below.

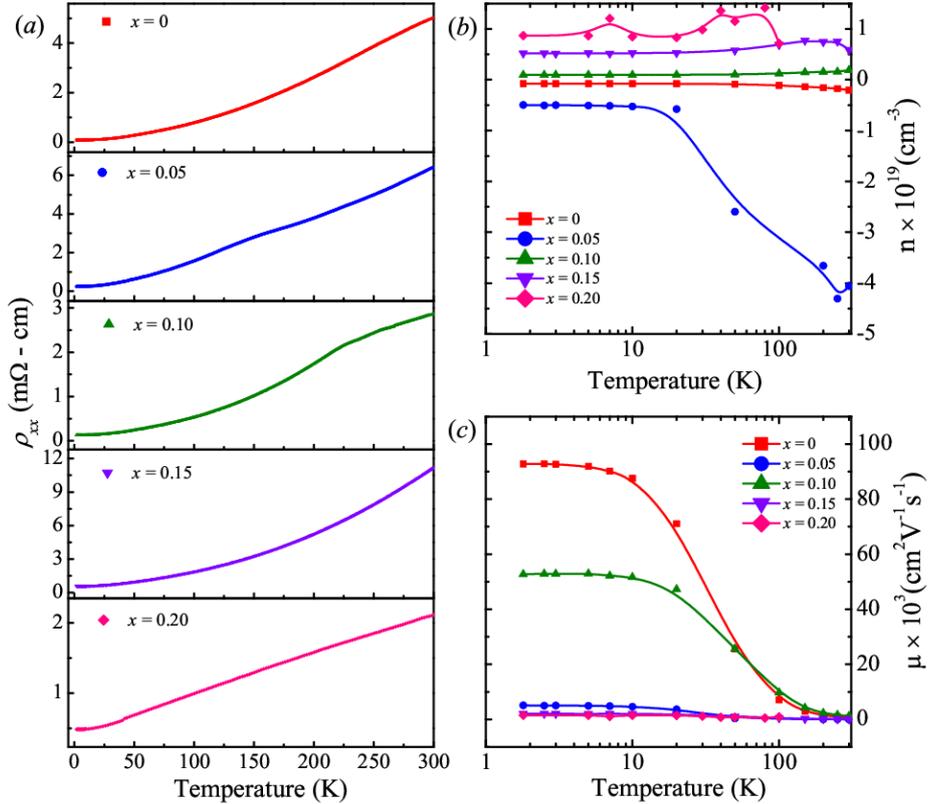

Figure 1: (a) Temperature dependence of zero-field longitudinal resistivity ($\rho_{xx}$) for $Pd_xBi_2Te_3$ (*x* = 0, 0.05, 0.10, 0.15, 0.20) of crystals, and (b), (c) show the variation of carrier density (*n*) and mobility (μ) with temperature, respectively, calculated from Hall resistivity data.

Figure 2(a) shows magnetoresistance (MR % = [$\rho_{xx}$(H) − $\rho_{xx}$(0)/$\rho_{xx}$(0)] × 100%) for the $Pd_xBi_2Te_3$ crystals with magnetic field perpendicular to the sample surface (*ab* -plane) as well as the current direction. The low temperature

($T$ = 1.8 K) MR values for $Bi_2Te_3$ at 8 T field is ~1500 % which is larger than previously reported value of MR = 540 % at 7 T by Shrestha et al.[26,27], MR = 400 % at 13 T by Wang et al.[28], MR = 1000 % at 10 T by Singh et al.[29], and MR = 267 % at 9 T by Barua et al.[30] With the increasing Pd content, MR value decreases significantly from 1500 % ($x$ = 0) to 300 % ($x$ = 0.20). As seen from the Fig. 2(a), the MR of $Pd_xBi_2Te_3$ showed a quadratic dependence on magnetic fields. Figure 2(b) shows that the MR value remains almost unchanged up to 10 K, then starts to decrease on further increase of temperature. At 200 K, the MR curves of all Pd concentration overlaps and have similar values at room temperature. The decrease in MR value with temperature could also be associated with the increase in carrier concentration. The MR curves show parabolic field dependence at higher temperatures as well as with the increase in Pd content (See Fig. S6 [16]). Generally, high MR values are realized in semimetals, where electron and hole pockets coexist at the Fermi level.[31] A large value of MR and high mobility have been a feature of semimetals such as Bismuth, graphene, zero-gap semiconductors silver chalcogenides ($Ag_{2+\delta}Se$, $Ag_{2+\delta}Te$), Dirac and Weyl materials, and Heusler compounds.[31-36] The linear band dispersion as revealed by ARPES may results in high mobility, possibly due to high Fermi velocity of massless particles.[37] Moreover, a large non-saturating MR is expected in such systems.

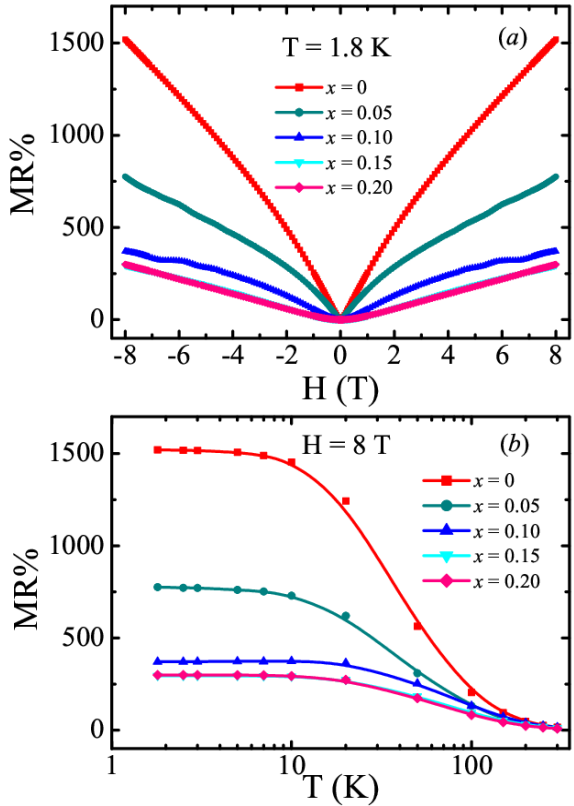

Figure 2: Magnetic field dependence of magnetoresistance (MR) at $T$ = 1.8 K (a), and temperature dependence of MR at 8 T (b) for $Pd_xBi_2Te_3$ ($x$ = 0 - 0.20).

Interestingly, the MR measured on the crystal of $Pd_{0.05}Bi_2Te_3$ with the broad hump shape anomaly around 100-150K in $\rho_{xx}(T)$ shows cusp-like feature at low magnetic field which is suggestive of WAL effect. To further examine the WAL effect, we show magnetoconductance of $Pd_{0.05}Bi_2Te_3$ in Fig. 3. The WAL is a quantum correction to conductivity resulting from both the strong spin-orbit coupling in the bulk as well as the spin-momentum locking in the TSSs of TIs.[38] The cusp like positive MR in low fields due to the WAL are observed up to 20 K and reduces at high temperature. The WAL effect is derived from the destructive interference between two time-reversed electron paths around an impurity or scattering point. This destructive interference reduces the elastic backscattering, thereby increasing the conductivity. We have analyzed the WAL data using the well-known Hikami-Larkin-Nagaoka (HLN) model[39]:

$$\Delta G_{xx}(H) = G_{xx}(H) - G_{xx}(0)$$
$$\Delta G_{xx}(H) = -\alpha \frac{e^2}{2\pi^2 \hbar}\left[\ln\frac{\hbar}{4Hel_\phi^2} - \psi\left(\frac{1}{2} + \frac{\hbar}{4Hel_\phi^2}\right)\right]$$

where, $l_\phi$ is the phase coherence length and $\psi(x)$ is the digamma function.

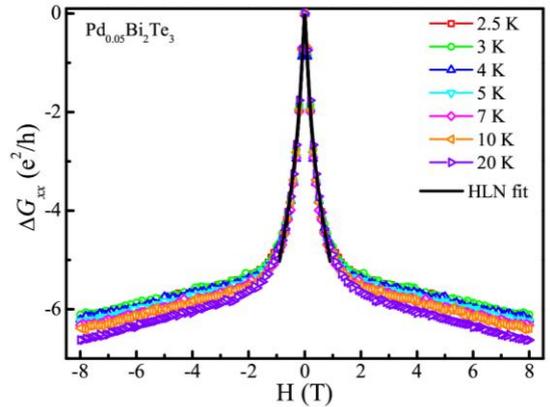

Figure 3: Weak antilocalization (WAL) effect in $Pd_{0.05}Bi_2Te_3$: the change in magnetoconductance ($\Delta G_{xx}$) at low temperatures and HLN fit (black line) in field range ±1 T.

The prefactor $\alpha$ is expected to be -0.5 for the 2D topological surface states of TIs. We have shown the fitting (black line) of the magnetoconductance curve in the range ±1 T in Fig. 3. The obtained $\alpha$ values (~ -7) at $T$ = 2.5 K is larger than the previously reported values in the range -0.4 to -2.5 for thin film as well as bulk samples of TIs.[40-44] Such large values can also be understood by assuming that some of the bulk channels also conduct together with the surface states of TIs. Additionally, the large value of magnetoconductance in this material is also consistent with the large value of $\alpha$. The $l_\phi$ values (~97 nm) are comparable to that discussed in the literature.[26]

In the field dependent resistivity of $Pd_{0.05}Bi_2Te_3$ and $Pd_{0.10}Bi_2Te_3$, we have observed Shubnikov-de Haas (SdH) oscillations. The quantum oscillation is an important probe to analyze the shape and sizes of the Fermi surface as well as topological properties of carriers. We have observed the oscillations at low temperatures (1.8 - 10 K) and magnetic field in the range of 4 to 8 T. The observed quantum oscillations in $\Delta\rho_{xx}$ for the $x = 0.05$ and $x = 0.10$ samples have been plotted in Figs. 4(a) and 4(c) as a function of inverse magnetic field at different temperatures ($\Delta\rho_{xx}$ has been determined by subtracting the polynomial background from $\rho_{xx}$). The magnitude of oscillations in $Pd_{0.10}Bi_2Te_3$ is larger than that in $Pd_{0.05}Bi_2Te_3$. The amplitude of SdH oscillations decreases with increase in temperature due to the thermal damping factor.

In Figs. 4(b) and 4(d) the fast Fourier transform (FFT) amplitudes for the subtracted oscillations are plotted with the frequency. The FFT amplitudes of both samples decrease on increasing temperature. The $x = 0.05$ sample shows a prominent peak at the oscillation frequency of $F = 2.75$ T for all temperatures whereas the $x = 0.10$ sample has a major peak at $F = 4$ T for 1.8 K which gradually shifts to 7.2 T for higher temperatures. The frequencies of these oscillations are proportional to the extremal cross-section of Fermi surface ($A_F = A(E_F)$) perpendicular to the applied magnetic field, based on the Onsager relation $F = \hbar/2\pi e\, A(E_F)$. The calculated values of Fermi surface cross-section assuming a circular Fermi surface ($A_F = \pi k_F^2$), Fermi wave vector ($k_F$), Fermi velocity $v_F = \hbar k_F/m_c$, and the electron mean free path ($l_F = v_F \tau$) are presented in Table 1. The frequencies of 2.75 T and 4 T imply a cross section of $2.68\times10^{12}$ cm$^{-2}$ and $3.83\times10^{12}$ cm$^{-2}$ respectively for $x = 0.05$ and 0.10 samples. These values are almost two orders less than what is measured from Hall resistivity: $4.9\times10^{18}$ cm$^{-3}$ for $x = 0.05$ and $9.5\times10^{17}$ cm$^{-3}$ for $x = 0.10$.

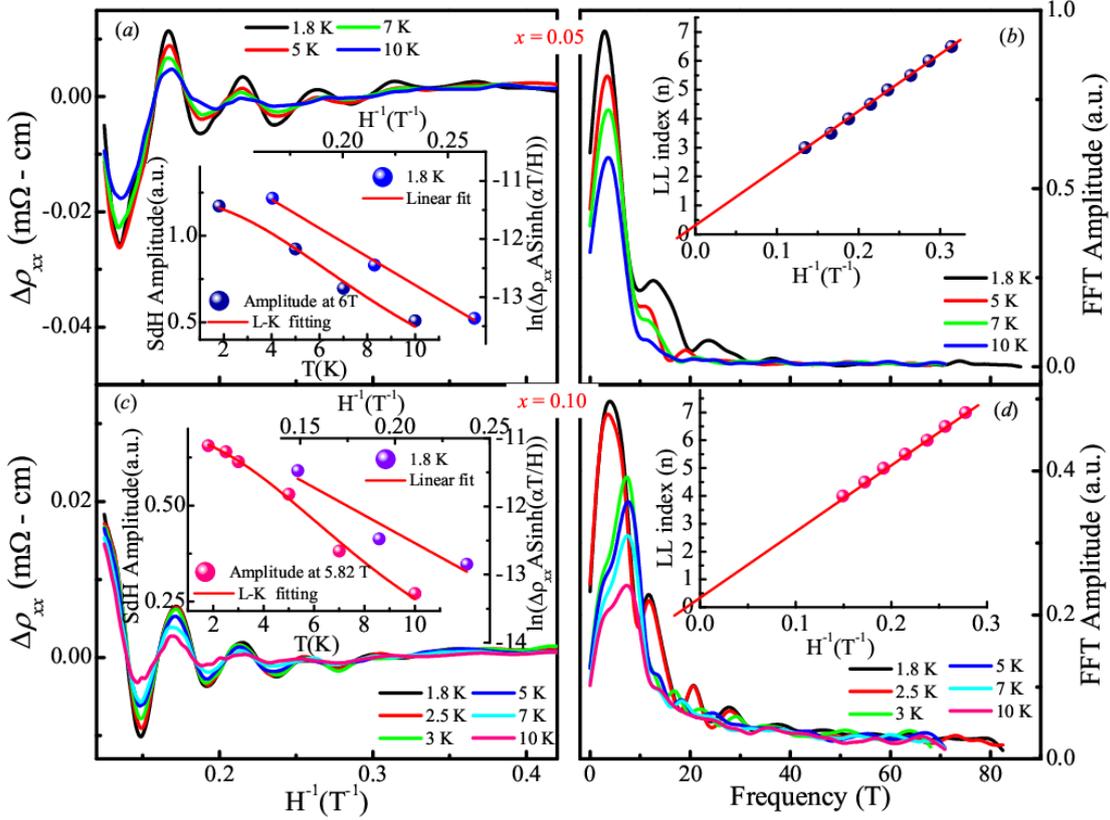

Figure 4:(a, c) Resistivity as a function of $1/H$, after subtracting the background contribution, inset shows the (Left) L-K formula fit and (Right) Dingle plot; (b, d) the fast Fourier transform analysis for $x = 0.05$ and $x = 0.10$ with fundamental frequency at 2.75 T and 4 T, respectively. Inset (Fig. (b), (d)) shows the Landau level fan diagram. The zeroth Landau level, $n_0$ is 0.33 and 0.35 for $x = 0.05, 0.10$, respectively.

We assume, therefore, significant contributions from the bulk valence bands to the Hall resistivity measurements.

The temperature and field dependent SdH oscillations amplitude in the resistivity ($\Delta\rho_{xx}$) were analyzed to yield transport parameters. According to the Lifshitz - Kosevich (L-K) expression, oscillation amplitude in $\Delta\rho_{xx}$ is given by

$$\Delta\rho_{xx} = \rho_T \rho_D$$

where $\rho_T = \dfrac{2\pi^2 k_B T}{\hbar\omega_c \sinh\dfrac{2\pi^2 k_B T}{\hbar\omega_c}}$, $\rho_D = e^{-\dfrac{2\pi^2 k_B T_D}{\hbar\omega_c}}$

Here, $\rho_T$ is the thermal damping factor and $\rho_D$ is the Dingle damping factor, $\omega_c = eH/m_c$ is the cyclotron frequency and $T_D$ is the Dingle temperature. In the Lifshitz-Kosevich theory, the thermal smearing of SdH oscillations is determined by the magnitude of cyclotron mass, $m_c$. As the temperature increases, the ratio of thermal energy to the cyclotron energy increases leading to the decrease in amplitude of oscillations. A fitting to the temperature dependent SdH oscillation amplitude shown in the inset of Fig. 4(a) and Fig. 4(c) at a fixed magnetic field $H$, yields the cyclotron mass $m_c$ of ~ $0.105 m_e$ for both $x = 0.05$ and $x = 0.10$. The effective mass obtained here are in fair agreement with that reported value ~ $0.09\ m_e$ for the undoped Bi$_2$Te$_3$.[27,29,30,45]

The Dingle temperature ($T_D$) is calculated from the field dependence of the SdH oscillation amplitude at fixed temperature. It yields the same field dependence as does the impurity scattering.[46] The Dingle plot is shown in the insets of Figs. 4(a), and 4(c), in which $\ln(\Delta\rho_{xx} A \sinh \alpha T/H)$ against $1/H$ is plotted, where $\alpha = 2\pi^2 k_B m_c/\hbar e$. A straight line fit to this plot generates a slope that contains Dingle temperature, given by $-2\pi^2 k_B m_c T_D/\hbar$. From $T_D$, quantum scattering lifetime $\tau$ for the carriers is calculated based on the relation $\dfrac{\hbar}{\tau} = 2\pi k_B T_D$. Then, we can further calculate surface carrier mobility ($\mu_s = e\tau/m_c = el^{SdH}/\hbar k_F$) and mean free path ($l^{SdH}$) of surface carrier. The obtained lower Dingle temperature value implies weaker impurity scattering that results in higher quantum mobility.[46] The calculated surface carrier mobility is less than that calculated from Hall measurement.

To further study the topological properties of the surface states, we estimated the Berry phase value from the Landau level fan diagram, that is given by $n = F/2\pi H + \beta$, where $n$ is the $n^{th}$ Landau level, $F$ is the SdH oscillation period, $H$ is the magnetic field. The inset of Fig. 4(b, d) shows the Landau level fan diagram plotted as the $1/H$ positions of maxima and minima in the background subtracted SdH oscillations in resistivity as a function of the integer values known as Landau indices. The minima in resistivity correspond to integer values $n$ and maxima to $n+1/2$. The zeroth Landau level can be extracted from the linear extrapolation of the $1/H = 0$, which gives $n_0 = 0.33$ and $0.35$ for $x = 0.05$ and $0.10$, respectively. The Berry phase $\varphi = 2\pi\beta$ is 0, $\beta = 0$, for the normal fermions, and Berry phase is $\pi$, $\beta = 0.5$ for the Dirac fermions with the linear dispersion relation. The extracted values of $n_0$ in our results are closer to 0.5, indicating that the oscillations arise from the topological surface states with a finite contribution from bulk states. Thus, the analysis of SdH oscillations provide evidence of non-zero Berry phase for $x = 0.05$ and $0.10$ samples, which indicates the existence of the TSSs.

Table I. Parameters obtained from SdH oscillation and fast Fourier transform (FFT) analysis.

| **Obtained parameters** | $x = 0.05$ | $x = 0.10$ | Bi$_2$Te$_3$ |
|---|---|---|---|
| Fermi momentum $k_F$ ($10^6$cm$^{-1}$) | 0.93 | 1.11 | 3.4 [ref.[27,30,45]] |
| Fermi cross section $A_F$ ($10^{13}$cm$^{-2}$) | 0.27 | 0.38 | 3.63 [ref. [45]] |
| Surface carrier concentration $n_s$ ($10^{11}$cm$^{-2}$) | 0.68 | 0.97 | 9.5 [ref. [27,45]] |
| Bulk carrier concentration $n_b$ ($10^{16}$cm$^{-3}$) | 2.67 | 4.56 | 109 [ref. [29]] |
| Fermi velocity $v_F$ ($10^7$cm s$^{-1}$) | 1.02 | 1.22 | 5.16 [ref. [45]] |
| Dingle temperature $T_D$ (K) | 9.29 | 9.9 | 16 [ref. [45]] |
| Carrier lifetime $\tau$ ($10^{-13}$s) | 1.31 | 1.22 | 0.56 [ref. [45]] |
| Mean free path $l^{SdH}$ (nm) | 13.3 | 14.9 | 28 [ref. [45]] |
| Mobility $\mu_s$ (cm$^2$V$^{-1}$s$^{-1}$) | 2191 | 2046 | 1417 [ref. [45]] |
| Cyclotron mass $m_c$ | $0.105\ m_e$ | $0.105\ m_e$ | $0.09$ [ref. [45]] |
| $k_F l$ | 1.23 | 1.65 | 6.5 [ref. [30]] |

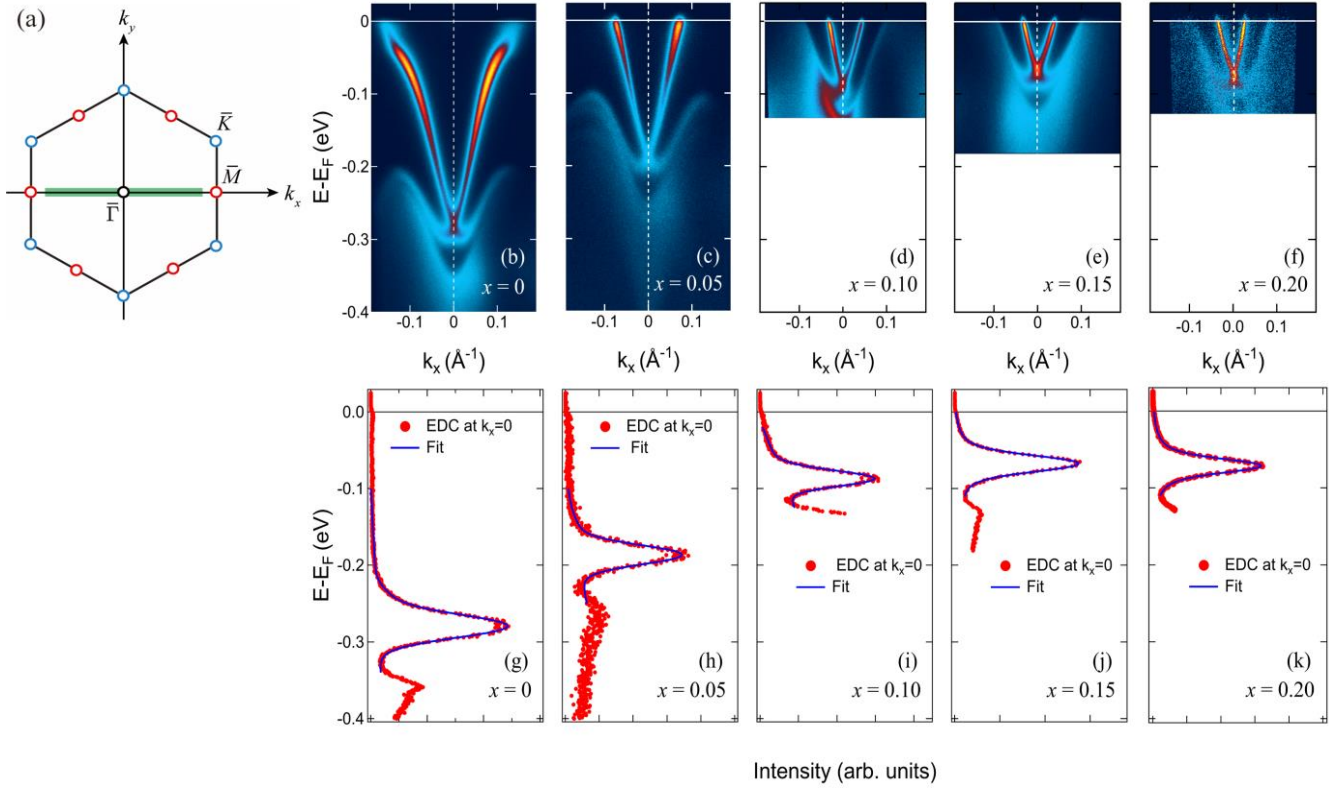

Figure 5: (a) Surface Brillouin zone (SBZ) with three high symmetry points, $\bar{\Gamma}$ (black), $\bar{M}$ (red), and $\bar{K}$ (blue). The ARPES spectra were measured along the green line, *i.e.*, along the $\bar{\Gamma} - \bar{M}$ high symmetry line in the SBZ. (b)-(f) The ARPES image plots of $Pd_xBi_2Te_3$ showing the rigid-band-like shift on Pd doping. (g)-(k) Energy distribution curves (EDCs) at the $\bar{\Gamma}$-point ($k_x = 0$). The EDC peak at the Dirac point can be fitted with a single Lorentzian suggesting no energy gap.

## B. High-resolution ARPES measurements

Figures 5(b)-5(f) represent the high-resolution ARPES image plots on $Pd_xBi_2Te_3$ ($0 \leq x \leq 0.20$) taken along the $\bar{\Gamma} - \bar{M}$ high symmetry direction (Fig. 5(a)). The energy distribution curves (EDCs) taken at the $\bar{\Gamma}$ point are shown in Figs. 5(g)-5(k). The EDC peaks at the Dirac point can be fitted by a single Lorentzian function suggesting the gapless TSS for $Pd_xBi_2Te_3$ ($0 \leq x \leq 0.20$) samples in agreement with non-zero Berry phase. One can see the Dirac cone like TSSs crossing the Fermi level as well as signals from the bulk valence bands projections. In the case of pristine $Bi_2Te_3$, there exist electron like conduction bands inside the TSSs for the upper part of the Dirac cone.[6,7] In this study, pristine $Bi_2Te_3$ is apparently *n*-type because the Fermi level intersects with the bulk conduction band (Fig. S7 [16]), the electron-like conduction band is compatible with the negative Hall coefficient in the bulk.

It is clearly seen from Figs. 5 (b)-5(f) that the initial addition of Pd into the crystal leads to hole doping of the entire electronic structure and both the TSSs and bulk derived valence states are shifted toward higher energy in a rigid-band-like manner. The energy shift continues up to $x = 0.1$ and stopped for $x \geq 0.10$. Note that bulk valence bands cross the Fermi level for the $x \geq 0.1$ samples, in agreement with the positive Hall coefficient in the bulk. It should be noted that the topological surface state of $x = 0.05$ is well-isolated from the bulk band at $E_F$, and the electric transport is governed by the surface state. On the other hand, for higher compositions, the bulk valence bands go across $E_F$, indicating that the contribution of the surface states might be masked by the bulk carriers in the transport properties. Note that other doped topological insulator $A_xBi_2Se_3$ (Sr, Nb, Cu) are reported to show downward shift of Dirac point with increase in dopant concentration.[47-50] The position and time dependent ARPES results on $x = 0.10$ and $x = 0.20$ are presented in supplemental information (See Fig. S8 and S9 [16]). We found the difference of the energy is at most ~8 meV. In the case of the pristine sample ($Bi_2Te_3$), the energy deviation was below the detection limit. These results indicate the magnitude of the homogeneity. After 24 hours of cleavage, the DP was shifted to lower energy likely due to the adsorption of residual gas on the sample surface.

In Figs. 6(a)-(c), on moving towards the lower binding energy (*i.e.,* from the $E_F$ to the Dirac point), the shape of constant energy contour for $Bi_2Te_3$ changes from a star-like to a circle with decreasing the cross-section area, whereas for Pd doped samples, it is already rounded hexagon at $E_F$.

Figure 6(d) shows the surface charge carrier density as a function of the Pd concentration. The initial Pd doping leads to a rapid reduction of carrier density of the TSS up to a value of $x = 0.1$, after that the TSS density remains essentially unchanged. The saturation of the doping level of the TSSs above $x = 0.1$ where the valence band maximum (VBM) starts to cross the Fermi level. One can expect a significant contribution from the bulk valence band for $x \geq 0.1$ sample.

In order to evaluate the $k_F$ value, we fit the momentum distribution curve (MDC) at the Fermi level using Voigt function (See Fig. S10 [16]).[51] As a result, we can estimate the charge carrier density $n_{2D,TSS} = \frac{\pi k_F^2}{(2\pi)^2}$ for the TSSs. The Fermi velocity ($v_F$) and effective mass ($m^*$) were determined by using the formula $v_F = \frac{1}{\hbar}\left(\frac{\partial E}{\partial k}\right)_{k=k_F}$ and $m^* = \frac{\hbar k_F}{v_F}$, respectively. We have also evaluated the linewidth $\Gamma = (\Delta E) = \left(\frac{\partial E}{\partial k}\right) \times (\delta k)$ and relaxation time $\tau = \frac{\hbar}{\Gamma}$ using the MDC width $\delta k$, which is a Lorentzian linewidth in the Voigt function (Gaussian line width represents the instrumental resolution).[51-53] It should be noted that ARPES linewidth at $E_F$ is sensitive to the lattice defects, and the observed linewidth for all our compositions are among the narrowest for the reported doped topological insulators.[6,18-24] Table II represents the parameters obtained for the TSSs only.

Table II: Parameters obtained from ARPES measurements along the $\bar{\Gamma} - \bar{M}$ high symmetry direction.

| Parameters | $x = 0$ | $x = 0.05$ | $x = 0.10$ |
|---|---|---|---|
| $k_F$ ($10^6$ cm$^{-1}$) | 14 | 7.2 | 4.2 |
| $A_F$ ($10^{13}$ cm$^{-2}$) | 62 | 16 | 5.5 |
| $n_{2D,TSS}$ ($10^{12}$ cm$^{-2}$) | 16 | 4.1 | 1.4 |
| $v_F$ ($10^7$ cm s$^{-1}$) | 3.1 | 4.0 | 3.1 |
| $m^*$ | $0.52 m_e$ | $0.21 m_e$ | $0.15 m_e$ |
| $\Gamma$ (meV) | --- | 16 | 4.1 |
| $\tau$ ($10^{-13}$ sec) | --- | 0.42 | 1.5 |
| MDC linewidth: $\delta k$ (Å$^{-1}$) | --- | 0.006 | 0.002 |
| Mean free path $l$ (nm) | --- | 17 | 50 |

## IV. DISCUSSION:

Now, we compare results obtained from the SdH oscillation and ARPES measurements. The Fermi wavenumber obtained from the SdH oscillation are almost identical for $x = 0.05$ and $0.10$ samples, $k_F \sim 1 \times 10^6$ cm$^{-1}$, and the values are smaller by a factor of 1/7 ($x = 0.05$) and 1/4 ($x = 0.10$) in comparison to that obtained from ARPES results. The reduction in the $k_F$ value obtained from magneto-transport is likely due to the band bending induced by the Schottky barrier leading to the shifting of the Fermi level towards the Dirac point. We assume the band bending is saturated if the valence band crosses the Fermi level as shown in Fig. 5. Based on the observed Fermi surfaces by ARPES, one can clearly see that the Fermi surface area for the Pd doped samples are reasonably approximated by $A_F = \pi k_F^2$. Since the area and surface carrier density is in proportion to $k_F^2$, the ARPES results yield 16 (= $4^2$) to 49 (= $7^2$) times larger values compared with the magneto-transport measurements. On the other hand, the cyclotron mass, $m_c \sim 0.1 m_e$ for $x = 0.05$ and $0.10$ samples, obtained from the SdH oscillation are close to the effective mass $m^* \sim 0.2 m_e$ evaluated in the ARPES experiments. We assume that the effective mass given by SdH and ARPES are similar in spite of the difference in the $k_F$ value because of the linearity of the band dispersion of the TSSs. Furthermore, we should note that the mean-free-path, $l = v_F \tau = 13-15$ nm, obtained from SdH oscillation are close to the values, $l = 1/\delta k = v_F \tau = 17 - 50$ nm, evaluated from ARPES. As the Berry phase is closer to $\pi$, we assume that the parameters obtained by the SdH oscillation mainly derive from the TSSs, while the carrier density is likely reduced due to the band bending.

It is important to note the difference in these two measurements. While the TSSs on the surface is directly observed by ARPES, the metal-semiconductor contacts are inevitable for the transport measurements. The Schottky barrier induces the band bending at the interface, which modifies the energy position of the Dirac point location in the near surface region.[18,54] Therefore, the Fermi wave number may be modified in the SdH oscillation measurements, and hence the surface carrier density. Furthermore, the SdH oscillation provide parameters averaged over the Fermi surface while ARPES results are along the $\bar{\Gamma} - \bar{M}$ high symmetry direction. We should note, however, the anisotropy of the Fermi surface shape is reduced for doped samples ($x > 0.05$). Furthermore, the ARPES signals are obtained from the beam spot size of < 10$\mu$m $\phi$, while SdH oscillation signals are averaged over ~1 mm$^2$ size samples.

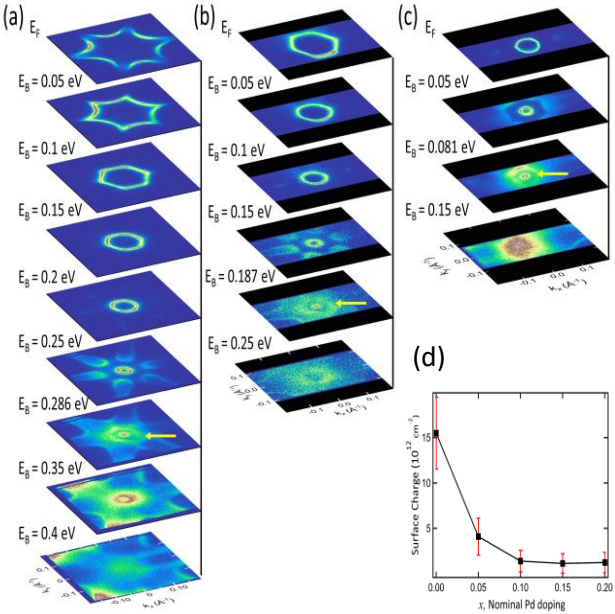

Figure 6. A set of constant energy contours at different energies (a) Bi$_2$Te$_3$, (b) Pd$_{0.05}$Bi$_2$Te$_3$ and (c) Pd$_{0.10}$Bi$_2$Te$_3$ respectively. Arrow indicates the position of the Dirac point. (d) The electron charge carrier density at the surface as a function of Pd doping concentration.

## V. CONCLUSION:

We have presented the effect of Pd doping on the magneto-transport and electronic properties of Bi$_2$Te$_3$ crystals. We have shown the evolution of various electronic parameters such as lifetime and mean-free-path of carriers at the Fermi level obtained through the ARPES and SdH oscillations studies. The observation of SdH oscillations, weak antilocalization, non-zero Berry phase in magneto-transport studies clearly indicated the contribution from the topologically nontrivial surface states. The Dirac cone-like surface states in ARPES confirms that Pd$_x$Bi$_2$Te$_3$ retains its topological properties upon Pd doping. Hall measurement shows the crossover from *n*-type charge carriers in undoped Bi$_2$Te$_3$ to *p*-type charge carriers upon Pd doping in the bulk. This is consistent with the ARPES results, in which both Dirac cone and bulk derived valence states are pushed towards the Fermi level upon Pd addition. Besides the energy shift of the Dirac point, the quasiparticle properties such as effective mass and mean-free-path are consistent with the SdH oscillation and ARPES measurements, which are consistent with the linearity of the band dispersion of the TSS.

## ACKNOWLEDGEMENT:

SS acknowledges IIT Mandi for HTRA fellowship. CSY acknowledges the experimental facility at AMRC, IIT Mandi. The ARPES measurements were performed with the approval of the Proposal Assessing Committee of the Hiroshima Synchrotron Radiation Center (Proposal Number: 19BU009). We thank N-BARD, Hiroshima University for supplying the liquid Helium.

#*Present address*: Experimentelle Physik VII, Universität Würzburg, Am Hubland, D-97074 Würzburg, Germany, EU, and Würzburg-Dresden Cluster of Excellence ct.qmat, Germany, EU.

# Supplemental Material

# Magneto-Transport and High-Resolution Angle-Resolved Photoelectron Spectroscopy Studies of Palladium Doped $Bi_2Te_3$

Shailja Sharma[1], Shiv Kumar[2], G.C. Tewari[3], Girish Sharma[1], Eike F. Schwier[2, #], Kenya Shimada[2], A. Taraphder[4], and C.S. Yadav[1]*

[1]*School of Basic Sciences, Indian Institute of Technology Mandi, Mandi-175075 (H.P.) India*
[2]*Hiroshima Synchrotron Radiation Center, Hiroshima University, Hiroshima, Japan*
[3]*Department of Chemistry and Material Science, Aalto University, FI-00076 Aalto, Finland*
[4]*Department of Physics, Indian Institute of Technology Kharagpur, Kharagpur (W.B) India*
*Email: shekhar@iitmandi.ac.in

Figure S1 (I) depicts the X-ray diffraction pattern on the crystals showing peaks corresponding to the (*00l*) planes, suggesting the crystals are grown along the *ab* plane. Figure S1 (II) shows the picture of crystals cleaved from the bulk ingot and rhombohedral crystal structure of $Pd_xBi_2Te_3$ drawn using *VESTA* software. The crystal structure shows the intercalated position of palladium atoms in the van der Waals gap between the quintuple layers (QLs). Figure S1 (III) shows the Laue diffraction pattern on the crystals.

Figure S2 (I) shows the Rietveld refined patterns of the powdered samples using the *FullProf* Suite, fitted with the rhombohedral structure ($R\bar{3}m$) of $Bi_2Te_3$. A comparison of the lattice parameters for all the compositions is tabulated in Table S1. The atomic radius of Pd (1.40 Å) is comparable to atomic radius of Bi (1.60 Å) than the Te (1.15Å). This may lead to the formation of secondary phase with increase in Pd content in $Bi_2Te_3$. The lattice parameters obtained in this study are indicative that the lattice expands with the Pd concentration. However, we find an impurity peak at $2\theta = 30.8°$ which corresponds to $Pd_9Te_4$.

Figure S2 (II, III) shows the FE-SEM images taken on crystals showing the layered morphology on different scales (5 and 10 µm).

Figure S3 shows the temperature dependence of longitudinal resistivity in zero-field for flakes (three for each composition) in addition to that presented in main manuscript.

Figure S4 presents the magnetization measurements results on $Pd_{0.20}Bi_2Te_3$ which reveal finite superconductivity (a) temperature dependence of magnetization, M(T) with superconducting transition ($T_c$) at 2.2 K, (b) magnetic field dependence of magnetization, M(H) plot at 1.8 K, showing superconducting hysteresis loop.

Figure S5 shows the magnetic field dependence of Hall resistivity at different temperatures (1.8 - 300 K) for the samples presented in main manuscript.

Figure S6 shows the magnetic field dependence of magnetoresistance (MR%) at different temperatures (1.8 – 300 K) for the samples presented in main manuscript.

Figure S7 shows Fermi surface of $Bi_2Se_3$ measured by ARPES. While the spectral intensity is weak, one can see spectral intensity of the bulk conduction band around the $\bar{\Gamma}$ point.

Figure S8(a) shows ARPES spectra of *x* = 0.10 from arbitrarily selected three different points just after cleaving (fresh surface), and Fig. S8(b) shows corresponding ARPES spectra after 24 hours of cleavage (aged surface). By the fitting of

EDC spectra at the $\bar{\Gamma}$ point, the Dirac point energies were evaluated to be -0.082 ~ -0.088 eV, which were consistent with our previous measurements -0.081 ~ -0.087 eV. The Dirac point energy may vary by ~10% depending on measurement points on the surface, which falls within the error bars. Note that the Dirac point energy was shifted to lower energy (electron doping) by 0.035-0.042 eV after 24 hours, which was likely due to the adsorption of the residual gas on the sample surface.

Figure S9(a) shows ARPES spectra of $x = 0.20$ from arbitrarily selected two different points just after cleaving, and Fig. S9(b) shows corresponding ARPES spectra after 24 hours of cleavage. We found that the Dirac point was -0.070 ~ -0.072 eV for fresh surface. After 24 hours, the Dirac point was shifted to lower energy by 0.078 - 0.102 eV due to the deterioration of the surface. The energy shifts for $x = 0.20$ were larger than those for $x = 0.10$ and hence more sensitive to the surface deterioration.

Figure S10 shows fitting results of EDCs at the $\bar{\Gamma}$ point. We could reproduce the EDC using single Lorentzian function, which indicates no gap opening at the Dirac point.

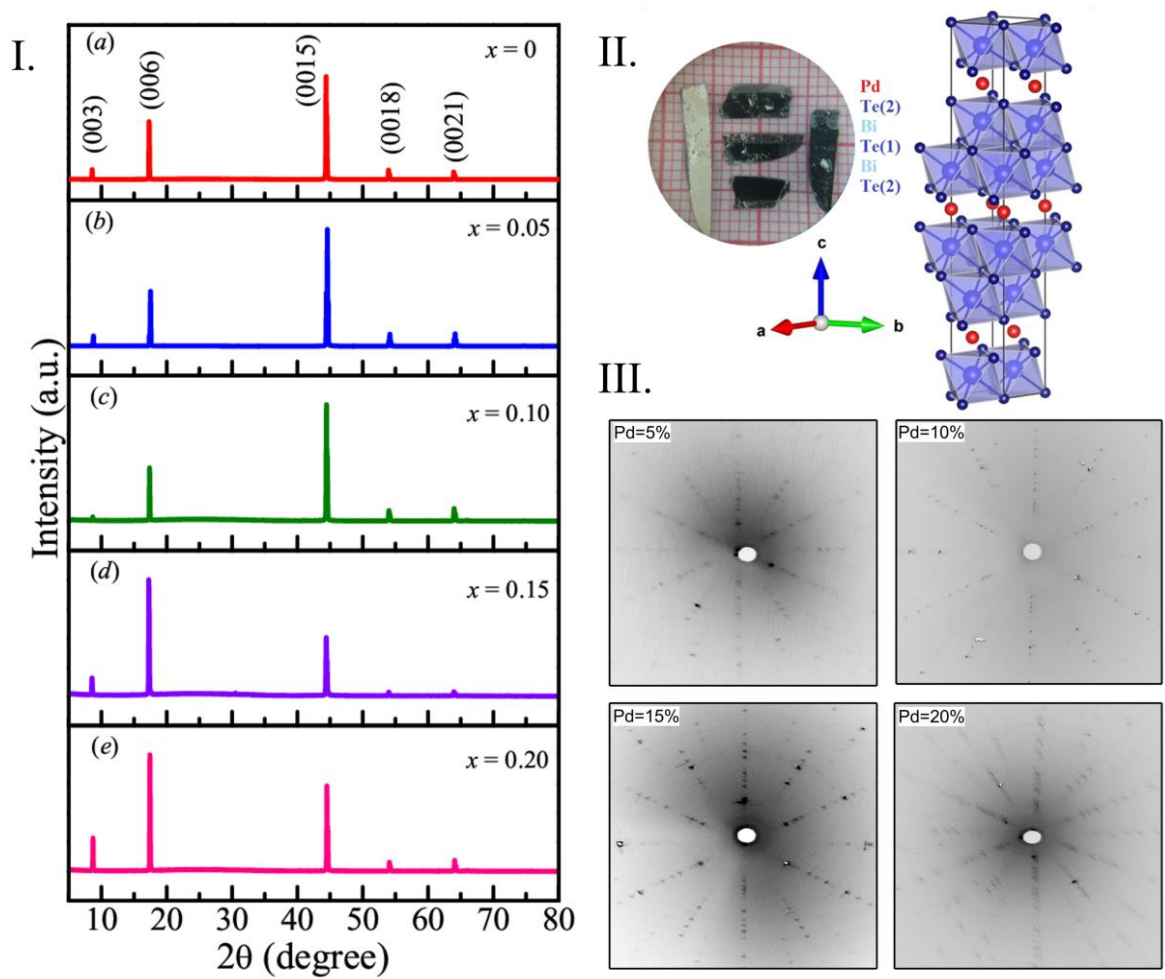

Figure S1:(I) Room temperature X-ray diffraction results on crystals of $Pd_xBi_2Te_3$ ($x = 0$, 0.05, 0.10, 0.15, 0.20) showing (00$l$) reflections only (Left), (II) Cleaved flakes from a few cm sizes grown crystal and crystal structure drawn using *VESTA* software (Right), (III) Laue diffraction pattern on crystals.

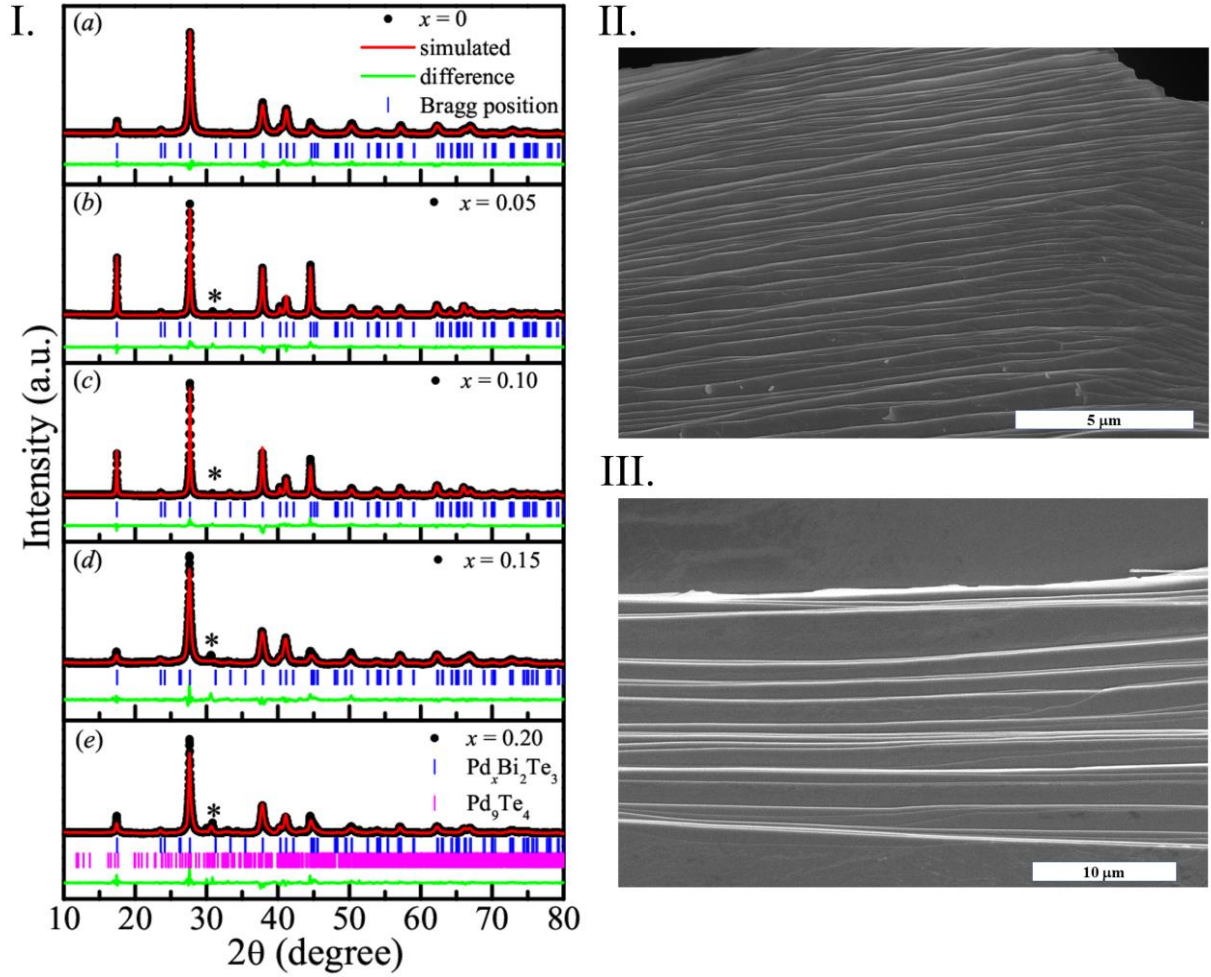

Figure S2: (I) Rietveld refined patterns of the powdered Pd$_x$Bi$_2$Te$_3$, (II, III) FE-SEM images showing the layered morphology of the crystal.

Table SI. Lattice parameters for Pd$_x$Bi$_2$Te$_3$ ($x$ = 0, 0.05, 0.10, 0.15, 0.20) after Rietveld refinement.

| $x$ | a = b (Å) | c (Å) | Volume (Å$^3$) | c/a |
|---|---|---|---|---|
| 0 | 4.3795(4) | 30.436(3) | 505.56(8) | 6.95 |
| 0.05 | 4.3845(2) | 30.4896(12) | 507.59(4) | 6.95 |
| 0.10 | 4.3827(2) | 30.4830(16) | 507.08(5) | 6.95 |
| 0.15 | 4.3895(5) | 30.410(4) | 507.42(11) | 6.93 |
| 0.20 | 4.3877(5) | 30.422(4) | 507.21(10) | 6.93 |

Table SII. The SEM-EDS elemental composition for Pd$_x$Bi$_2$Te$_3$ ($x$ = 0, 0.05, 0.10, 0.15, 0.20).

| Sample | Pd (at. %) | Bi (at. %) | Te (at. %) | Calculated composition |
|---|---|---|---|---|
| Bi$_2$Te$_3$ | 0 | 39.34 | 60.66 | Bi$_{1.95}$Te$_{3.00}$ |
| Pd$_{0.05}$Bi$_2$Te$_3$ | 0.68 | 38.92 | 60.40 | Pd$_{0.0338}$Bi$_{1.93}$Te$_{3.00}$ |
| Pd$_{0.10}$Bi$_2$Te$_3$ | 0.80 | 40.02 | 59.00 | Pd$_{0.0407}$Bi$_{2.03}$Te$_{3.00}$ |
| Pd$_{0.15}$Bi$_2$Te$_3$ | 0.83 | 37.13 | 62.04 | Pd$_{0.0401}$Bi$_{1.80}$Te$_{3.00}$ |
| Pd$_{0.20}$Bi$_2$Te$_3$ | 0.78 | 38.69 | 60.53 | Pd$_{0.0387}$Bi$_{1.91}$Te$_{3.00}$ |

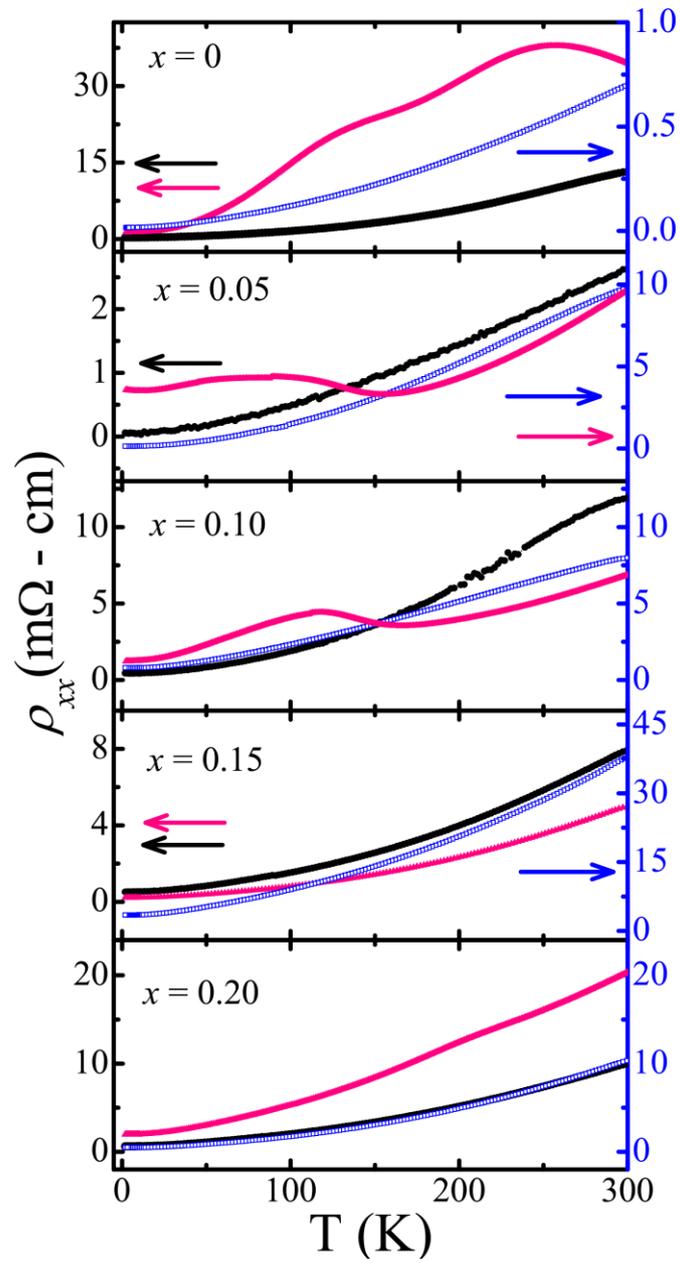

Figure S3: Temperature dependence of zero-field longitudinal resistivity on different flakes of Pd$_x$Bi$_2$Te$_3$ ($x$ = 0, 0.05, 0.10, 0.15, 0.20) obtained from the same quartz ampoule and same batch of crystals.

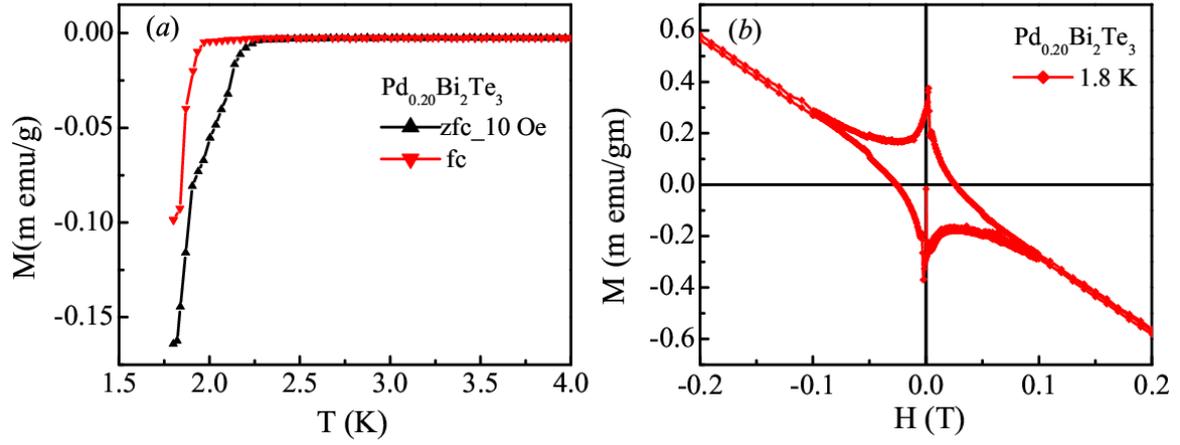

Figure S4: (a) M(T) plot at $H$ = 10 Oe (b) M(H) plot at $T$ = 1.8 K for $Pd_{0.20}Bi_2Te_3$.

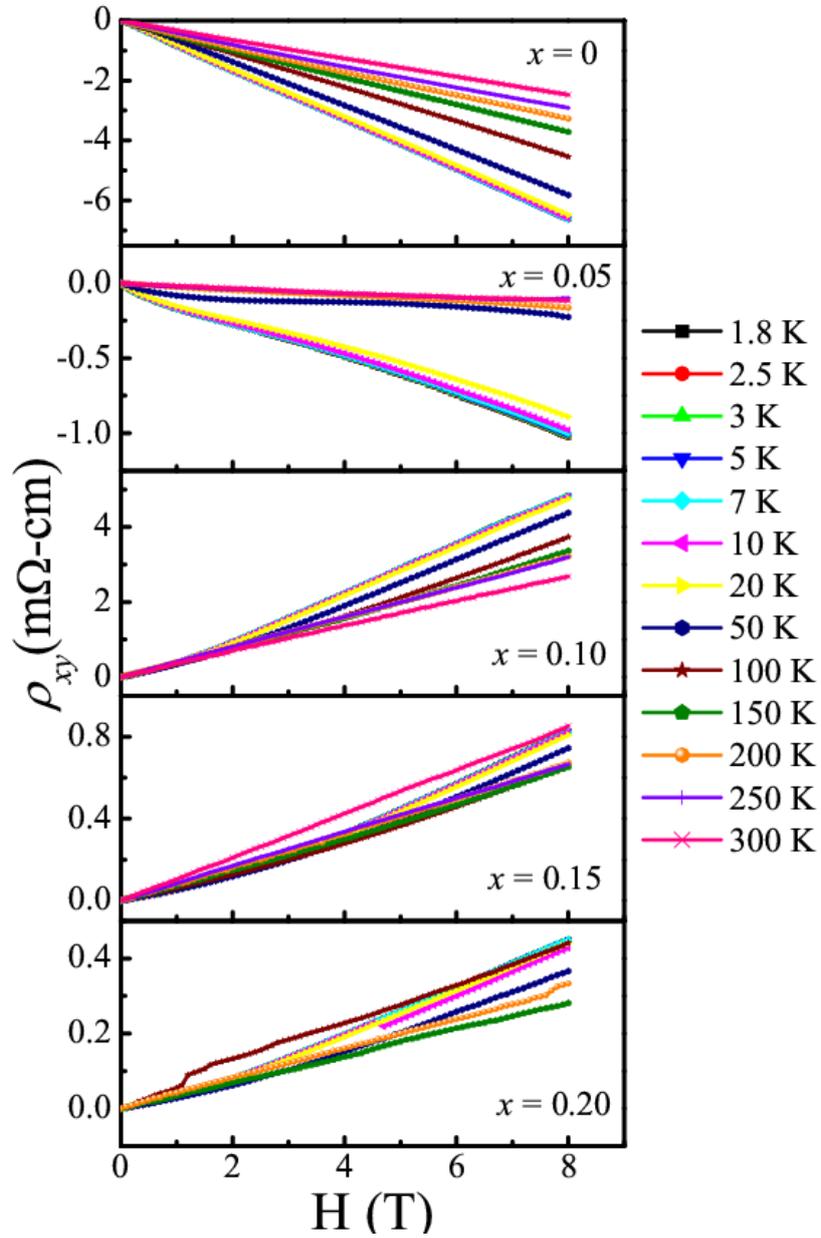

Figure S5: Hall resistivity ($\rho_{xy}$) plots with magnetic field (H) at various temperatures ranging from 1.8-300 K for $Pd_xBi_2Te_3$.

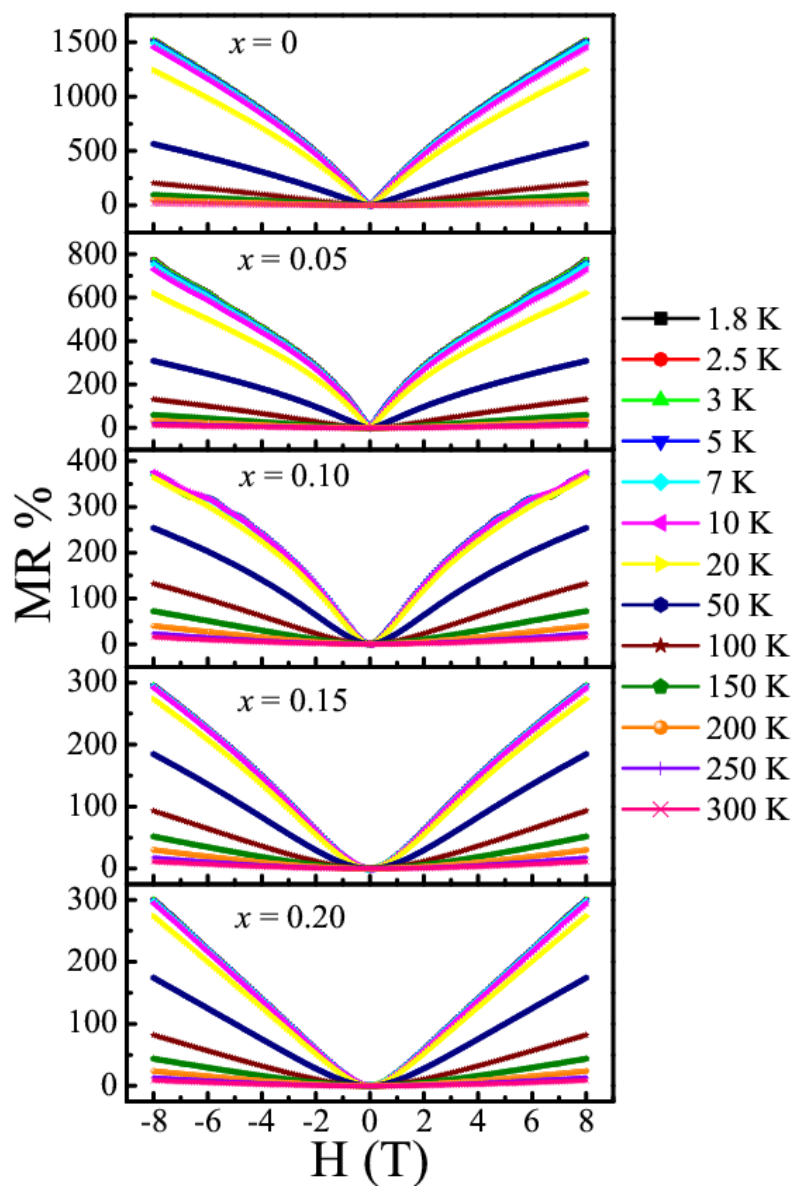

Figure S6: Magnetoresistance (MR %) with applied magnetic field (H) at various temperatures ranging from 1.8-300 K for $Pd_xBi_2Te_3$.

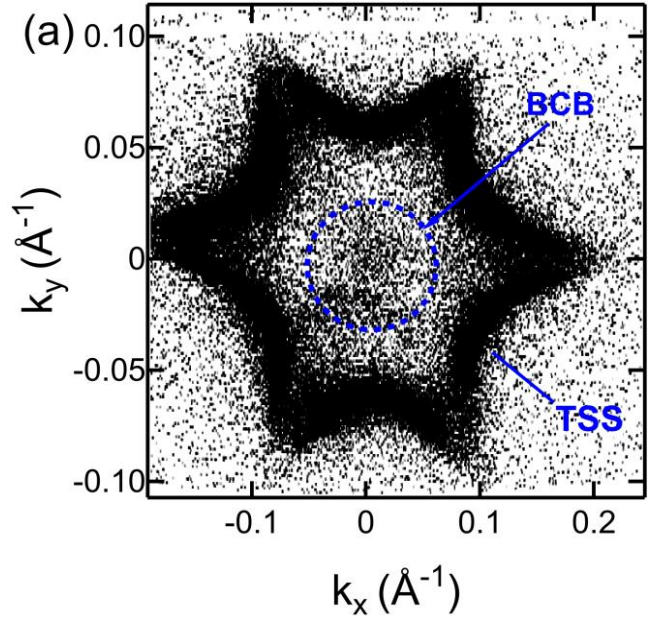

Figure S7: Fermi surface of $Bi_2Te_3$. The bulk conduction band crossing the Fermi level is discernible inside blue dashed circle.

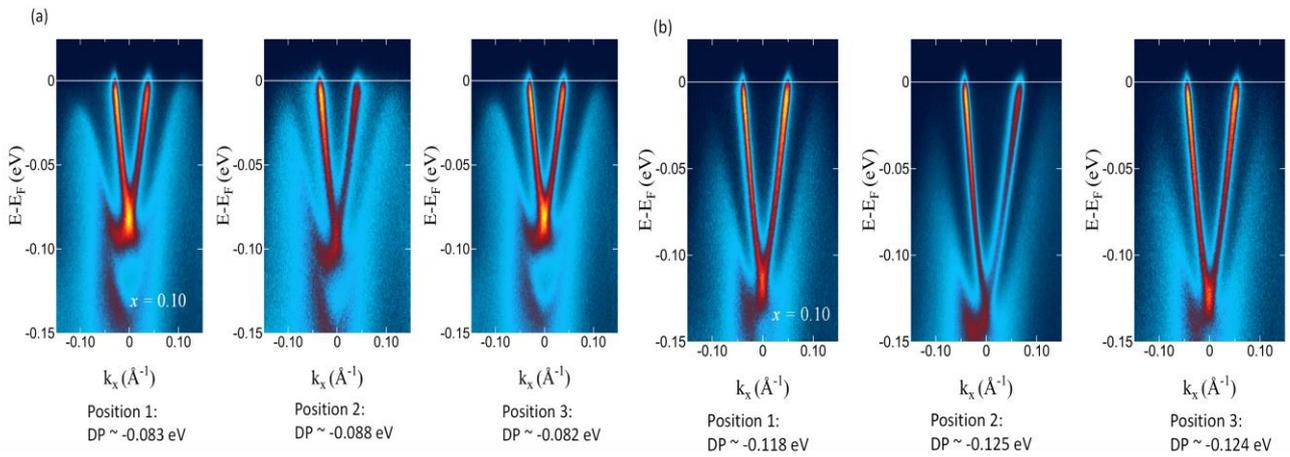

Figure S8: (a) ARPES measurement for $x = 0.10$ at three different positions on the sample surface along the $\bar{\Gamma} - \bar{M}$ high symmetry line in the SBZ. (b) ARPES measurement of the sample after 24 hrs at same positions.

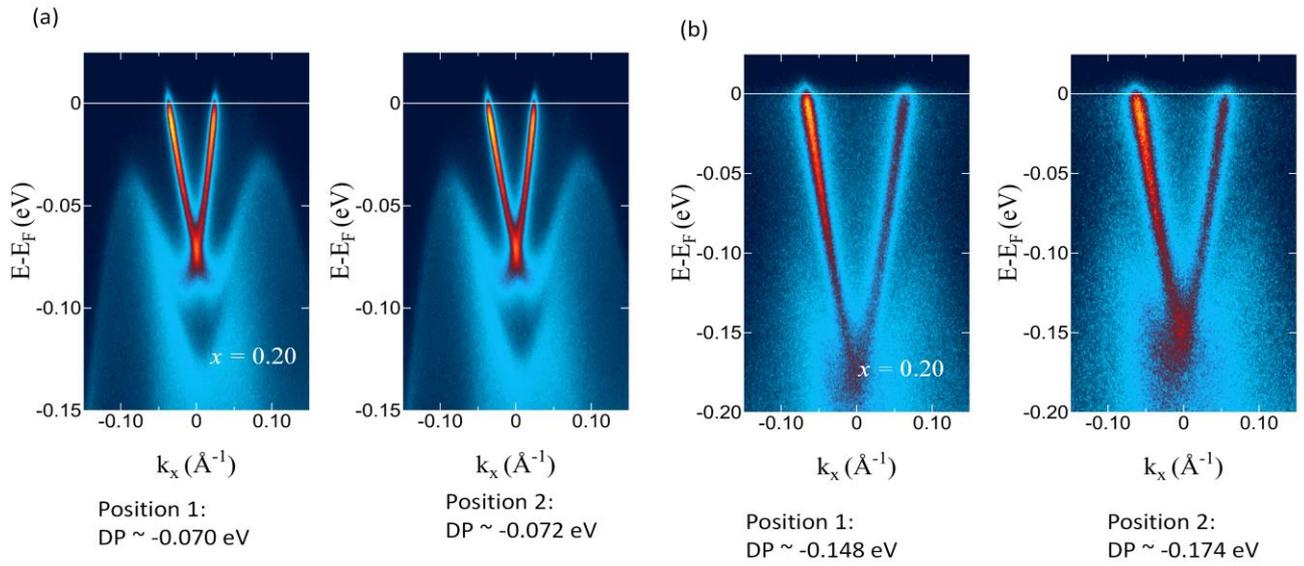

Figure S9: (a) ARPES measurement for $x = 0.20$ at two different positions on the sample surface along the $\bar{\Gamma} - \bar{M}$ high symmetry line in the SBZ. (b) ARPES measurement of the sample after 24 hrs at same positions.

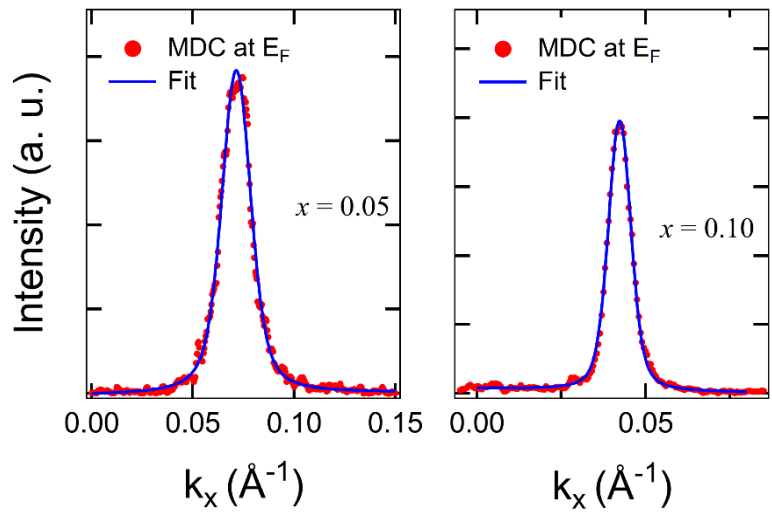

Figure S10: The MDC fitting at the Fermi level using Voigt function.